\documentclass[10pt,floatfix,twocolumn,superscriptaddress,amsmath,amssymb,aps,prd]{revtex4-2}
\usepackage[utf8]{inputenc}
\usepackage{siunitx}
\usepackage{hyperref}
\providecommand{\orcidlink}[1]{\hspace{1mm}\href{https://orcid.org/#1}{\textsuperscript{\tiny ID}}}
\usepackage{amsmath}
\usepackage{makecell}
\usepackage{array}
\usepackage{booktabs}
\usepackage{tabularx}
\usepackage[dvipsnames]{xcolor}
\usepackage{graphicx}
\usepackage[caption=false]{subfig}
\usepackage{microtype} 
\usepackage{multirow}
\usepackage[normalem]{ulem} 
 
\newcommand{\tdli}{State Key Laboratory of Dark Matter Physics, Key Laboratory for Particle Astrophysics and Cosmology (MoE), Shanghai Key Laboratory for Particle Physics and Cosmology, New Cornerstone Science Laboratory, Tsung-Dao Lee Institute, Shanghai Jiao Tong University, Shanghai 201210, China}
\def\sjtuphys{State Key Laboratory of Dark Matter Physics, Key Laboratory for Particle Astrophysics and Cosmology (MoE), Shanghai Key Laboratory for Particle Physics and Cosmology, School of Physics and Astronomy, Shanghai Jiao Tong University, Shanghai 200240, China}
\def\MESJTU{School of Mechanical Engineering, Shanghai Jiao Tong University, Shanghai 200240, China}
\def\SPEIT{SJTU Paris Elite Institute of Technology, Shanghai Jiao Tong University, Shanghai 200240, China}
\def\SJTUSC{Shanghai Jiao Tong University Sichuan Research Institute, Chengdu 610213, China}

\def\BUAA{School of Physics, Beihang University, Beijing 102206, China}
\def\BUAACenter{Peng Huanwu Collaborative Center for Research and Education, Beihang University, Beijing 100191, China}
\def\BUAALab{International Research Center for Nuclei and Particles in the Cosmos \& Beijing Key Laboratory of Advanced Nuclear Materials and Physics, Beihang University, Beijing 100191, China}
\def\SCNT{Southern Center for Nuclear-Science Theory (SCNT), Institute of Modern Physics, Chinese Academy of Sciences, Huizhou 516000, China}

\def\USTClab{State Key Laboratory of Particle Detection and Electronics, University of Science and Technology of China, Hefei 230026, China}
\def\USTCdep{Department of Modern Physics, University of Science and Technology of China, Hefei 230026, China}

\def\YaLongSD{Yalong River Hydropower Development Company, Ltd., 288 Shuanglin Road, Chengdu 610051, China}
\def\scKeyLab{Jinping Deep Underground Frontier Science and Dark Matter Key Laboratory of Sichuan Province, Liangshan 615000, China}

\def\pku{School of Physics, Peking University, Beijing 100871, China}

\def\SDUdep{Research Center for Particle Science and Technology, Institute of Frontier and Interdisciplinary Science, Shandong University, Qingdao 266237, China}
\def\SDUlab{Key Laboratory of Particle Physics and Particle Irradiation of the Ministry of Education, Shandong University, Qingdao 266237, China}

\def\SYU{School of Physics, Sun Yat-Sen University, Guangzhou 510275, China}
\def\SYUSFI{Sino-French Institute of Nuclear Engineering and Technology, Sun Yat-Sen University, Zhuhai 519082, China}
\def\SYUzhuhai{School of Physics and Astronomy, Sun Yat-Sen University, Zhuhai 519082, China}
\def\SYUshenzhen{School of Science, Sun Yat-Sen University, Shenzhen 518107, China}

\def\NKU{School of Physics, Nankai University, Tianjin 300071, China}
\def\YTU{Department of Physics, Yantai University, Yantai 264005, China}
\def\FDU{Key Laboratory of Nuclear Physics and Ion-beam Application (MOE), Institute of Modern Physics, Fudan University, Shanghai 200433, China}
\def\CDUT{College of Nuclear Technology and Automation Engineering, Chengdu University of Technology, Chengdu 610059, China}
\usepackage{hyperref}
\begin{document}
\title{Search for Diffuse Supernova Neutrino Background of All Flavors with PandaX-4T}
\author{Yingxin Zhang}\affiliation{\SDUdep}\affiliation{\SDUlab} 
\author{Peiyuan Chen\orcidlink{0009-0008-8703-4495}}\affiliation{\sjtuphys}
\author{Wei Chen\orcidlink{0009-0009-5911-7135}}\affiliation{\sjtuphys}
\author{Xiaohua Chen}\affiliation{\tdli}
\author{Xun Chen\orcidlink{0000-0001-7961-7908}}\affiliation{\tdli}\affiliation{\SJTUSC}\affiliation{\scKeyLab}
\author{Yunhua Chen}\affiliation{\YaLongSD}\affiliation{\scKeyLab}
\author{Chen Cheng\orcidlink{0000-0003-0164-7538}}\affiliation{\BUAA}
\author{Xiangyi Cui}\affiliation{\tdli}
\author{Yuxin Cui}\affiliation{\BUAA}
\author{Manna Deng}\affiliation{\SYUSFI}
% ...
\collaboration{PandaX Collaboration}
\noaffiliation

\begin{abstract}
We present the first search for the diffuse supernova neutrino background utilizing coherent elastic neutrino-nucleus scattering with the PandaX-4T liquid xenon detector at the China Jinping Underground Laboratory. Based on a total exposure of $1.54~\mathrm{tonne \cdot year}$ from the commissioning and first science runs, we observe no significant event excess over the expected background. We therefore set an upper limit of $(2.27 - 2.43) \times 10^3~\mathrm{cm^{-2}s^{-1}}$ per flavor at 90\% confidence level on the $\nu_x$ flux (each of $\nu_\mu, \bar{\nu}_\mu, \nu_\tau, \bar{\nu}_\tau$) of the diffuse supernova neutrino background for incident neutrino energies above 19.3~MeV. The result is comparable to the current world best limit.  
\end{abstract} 

\maketitle

\textit{Introduction}. 
Massive stars ($M \gtrsim 8 M_\odot$) end their stellar evolution in core-collapse supernovae (CCSNe), releasing approximately 99\% of the gravitational binding energy ($\sim 10^{53}$ erg) of the proto-neutron star as neutrinos~\cite{Santos:2025plx, Jerkstrand:2025bea}. These neutrinos serve as a powerful tool for investigating the dynamics of stellar collapse, the nucleosynthesis of heavy elements, and the intrinsic characteristics of neutrinos. The extreme rarity of galactic CCSNe, occurring at a rate of fewer than three events per century, severely limits the observation of individual supernova explosions. To date, the only direct observations of neutrinos from a core-collapse supernova remain those from SN1987A, which were limited to electron antineutrinos~\cite{20230604, Kamiokande-II:1987idp}. The diffuse supernova neutrino background (DSNB)~\cite{Santos:2025plx, Beacom:2010kk, Suliga}, in contrast, represents an isotropic neutrino flux accumulated from all core-collapse supernovae throughout cosmic history, offering a more comprehensive probe of the stellar collapse population, without the need to wait for a rare, once-in-a-lifetime galactic event. 

At the Neutrino 2026 conference, the Super-Kamiokande (SK) collaboration reported a 2.6 $\sigma$ indication for the $\bar{\nu}_{e}$ component of the DSNB~\cite{SKneu2026}. Looking ahead, the JUNO experiment is expected to reach a 3 $\sigma$ discovery potential for the same signal within three years of data taking~\cite{Cheng:2021ais,Abusleme_2022}. For the $\nu_{e}$,  the Sudbury Neutrino Observatory (SNO) provided the best limit of $\Phi_{\nu_{e}} < 19~cm^{-2}  s^{-1}$ (90\% C.L.) within the energy range of 22.9–36.9 MeV \cite{PhysRevD.102.062006}. And the best constraints on the $\nu_{x}$ ($\nu_{\mu}$, $\nu_{\tau}$, $\bar{\nu}_{\mu}$, $\bar{\nu}_{\tau}$) component of the DSNB flux remain weak, with SK's neutrino-electron elastic scattering (ES) analysis yielding an upper limit of $\sim$ \SI{e3}{cm^{-2}.s^{-1}} at 90\% C.L for neutrino energies above 19.3~MeV~\cite{Lunardini_2008}. However, it should be noted that probing the $\nu_{x}$ channel is essential for confirming the standard scenario of supernova bursts and for discovering new physics beyond conventional mechanisms~\cite{Nunokawa:1997ct, Fischer:2020xjl}. 

The coherent elastic neutrino-nucleus scattering (CE$\nu$NS) channel provides a powerful avenue to probe the $\nu_x$ component of the DSNB. This channel requires sufficiently low momentum transfer to enable coherent scattering of neutrinos off all constituent nucleons within the nucleus. This coherence condition drastically enhances the scattering cross section and ensures equal sensitivity to all neutrino flavors. Meanwhile, direct-detection dark matter experiments are rapidly advancing in their sensitivity to low-energy nuclear recoils (NRs), opening new opportunities for CE$\nu$NS measurements~\cite{PANDA-X:2025eka, Pang:2024bmg,PRL_B8,jvqf-njpj,XENON:2026smt,XENON_2024ijk}. Lately, the LZ collaboration has submitted a preprint on the DSNB search via CE$\nu$NS~\cite{Xia:2024ysq}. 

In this paper, we report the first search for DSNB via CE$\nu$NS with the PandaX-4T~\cite{PandaX:2018wtu} liquid xenon dark matter detector at the China Jinping Underground Laboratory (CJPL). This analysis combines data from the commissioning run (Run0, 95~days) and the first science run (Run1, 164~days), corresponding to a total exposure of  1.54~tonne$\cdot$year~\cite{PhysRevLett2024}.  



\textit{Signal prediction}. 
The differential DSNB flux, $d\Phi/dE_\nu$, arriving at Earth is evaluated as a line-of-sight integral for past core-collapse events over cosmic history~\cite{2021ApJ...909..169K}:
\begin{equation}
\frac{d\Phi}{dE_\nu} = c \int \frac{d{N}_{cc}}{d{E_\nu}^{'}}\frac{d{E_\nu}^{'}}{dE_\nu}{{R}_{cc}}(z){\left| {\frac{d{t}_{c}}{dz}} \right|} \,dz, 
\end{equation}
where $c$ is the speed of light and the source term $dN_{cc}/dE_\nu'$ denotes the initial mass function (IMF)-weighted average neutrino spectrum per core-collapse event. $z$ is the redshift and  the relation $E_\nu'=(1+z)E_\nu$ defines the redshift-corrected neutrino energy at emission, which introduces the transformation factor $dE_\nu'/dE_\nu$. The function $R_{cc}(z)$ represents the comoving core-collapse rate density over cosmic history.  Finally, $|dt_c/dz| = [H_0(1+z)\sqrt{\Omega_m(1+z)^3+\Omega_\Lambda}]^{-1}$ dictates the cosmological conversion between redshift $z$ and cosmic time $t_c$ based on the Friedmann equation.

\begin{figure}[htbp]
    \centering
    \begin{minipage}{0.48\textwidth}
        \centering
        \includegraphics[width=\textwidth]{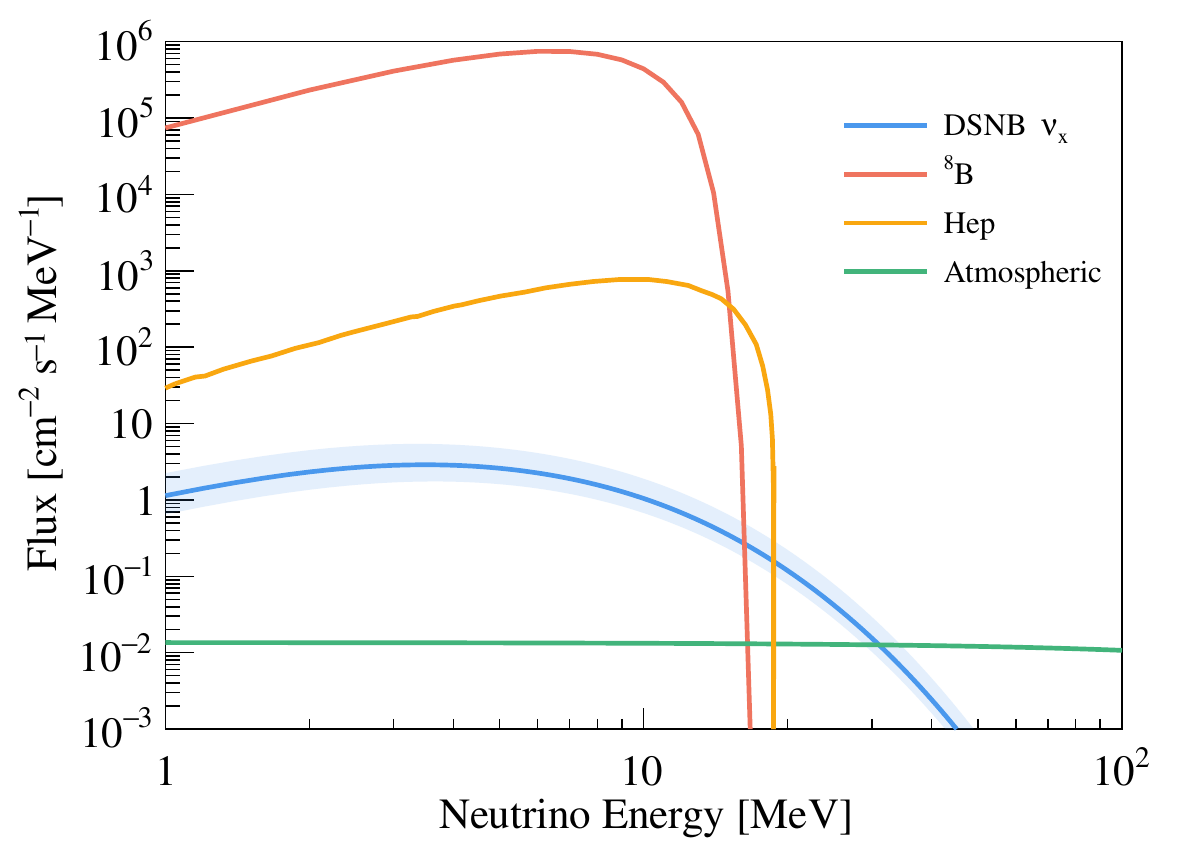} 
    \end{minipage}
    \hfill
    \begin{minipage}{0.48\textwidth}
        \centering
        \includegraphics[width=\textwidth]{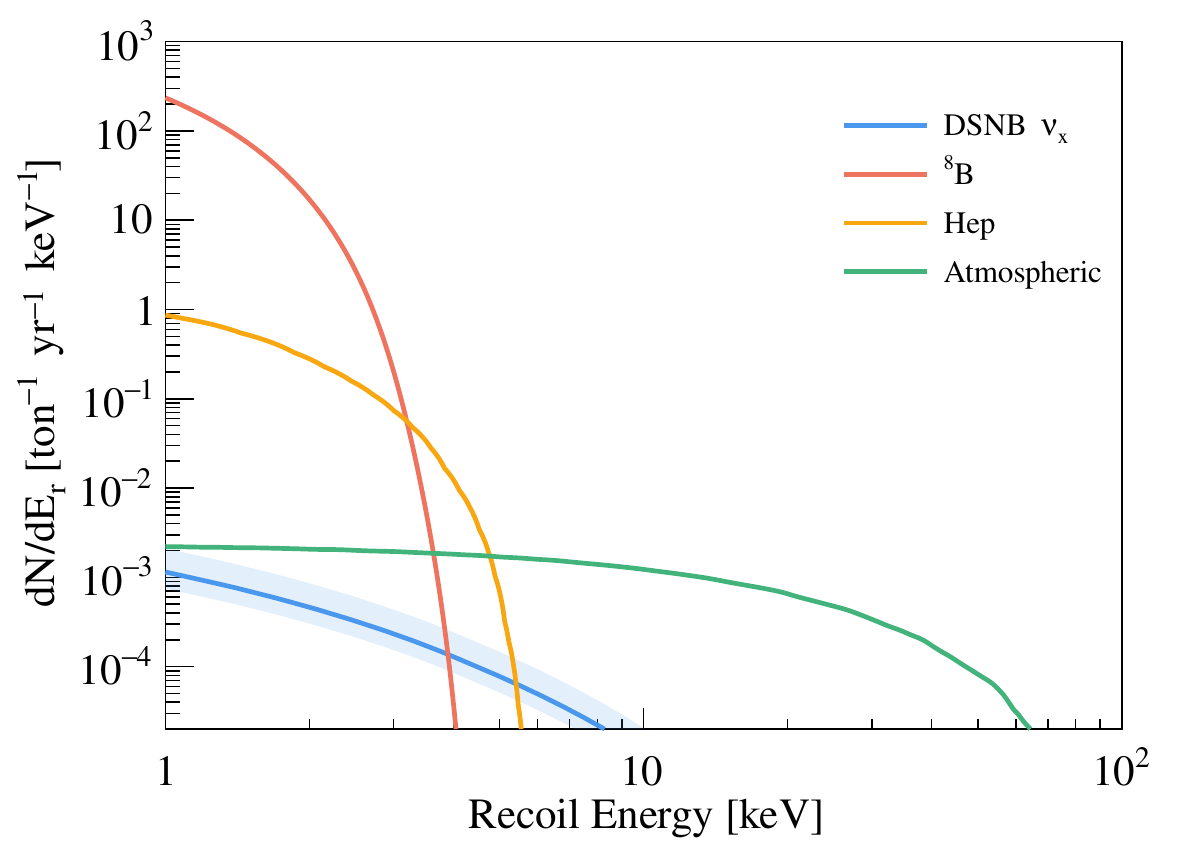} 
    \end{minipage}
    
    \caption{The DSNB $\nu_x$ (each of $\nu_\mu, \bar{\nu}_\mu, \nu_\tau, \bar{\nu}_\tau$) and the irreducible neutrino backgrounds~\cite{PRL_B8,PRD_NB,CJPL_atmospheric}. The top panel displays the incident neutrino flux spectra, while the bottom panel shows the expected CE$\nu$NS event rate as a function of NR energy in the liquid xenon target. The shaded band represents the uncertainty of the DSNB models.}
    \label{fig:flux_and_rate}
\end{figure}

The DSNB flux is evaluated assuming a standard $\Lambda$CDM cosmology and a comoving core-collapse rate density, $R_{cc}(z)$, inferred from the cosmic star-formation history. The neutrino emission is modeled using one-dimensional neutrino-radiation-hydrodynamic simulations spanning a discrete set of zero-age main-sequence (ZAMS) masses from $8.7\,M_\odot$ to $125\,M_\odot$, weighted by a modified Salpeter-A initial mass function (IMF). The time-integrated neutrino spectra from the Garching core-collapse supernova archive~\cite{Kresse2021CCSN} are used to construct the DSNB flux spectrum, with the uncertainty band reflecting the variation among different core-collapse progenitor models and their predicted neutrino emissions, as shown in Fig.~\ref{fig:flux_and_rate}. For clarity, only the $\nu_x$ spectrum, representing each of $\nu_\mu$, $\bar{\nu}_\mu$, $\nu_\tau$, and $\bar{\nu}_\tau$, is shown, as the $\nu_e$ component is more tightly constrained and a $2.6~\sigma$ indication of the $\bar{\nu}_e$ component has been reported. 


In the CE$\nu$NS process, a neutrino coherently and elastically scatters off a nucleus, inducing a NR:
\begin{equation}
\nu + A(Z, N) \rightarrow \nu + A(Z, N),
\end{equation}
where $A$ represents the mass number, $Z$ is the atomic number, and $N$ is the neutron number. For a given incident neutrino energy $E_\nu$, the differential CE$\nu$NS cross section with respect to the NR energy $E_r$ is expressed as~\cite{Suliga}

\begin{equation}
\frac{\mathrm{d}\sigma_{\nu A}}{\mathrm{d}E_r} = \frac{G_F^2}{4\pi} Q_W^2 \, m_A \left(1 - \frac{m_A E_r}{2 E_\nu^2} \right) \, F^2(q),
\label{eq:cross_section}
\end{equation}
where $G_F$ is the Fermi constant, $m_A$ is the nuclear mass, $Q_W = N - Z(1 - 4\sin^2\theta_W)$ is the weak vector nuclear charge, and $F(q)$ is the Helm-type nuclear form factor. The maximum NR energy is expressed by $E_{\text{max}} = \frac{2E_\nu^2}{m_A + 2E_\nu} \simeq \frac{2E_\nu^2}{m_A}$, corresponding to a scattering angle of $\theta = \pi$.

The expected differential NR event rate with respect to the NR energy per unit detector mass is then given by

\begin{equation}
\frac{\mathrm{d}R}{\mathrm{d}E_r} = \frac{N_t}{M} \int_{E_\nu^{\min}}^\infty \frac{\mathrm{d}\Phi}{\mathrm{d}E_\nu} \, \frac{\mathrm{d}\sigma_{\nu A}}{\mathrm{d}E_r} \, \mathrm{d}E_\nu,
\label{eq:event_rate}
\end{equation}
where $N_t/M$ is the number of xenon nuclei per unit mass, and $E_\nu^{\min} = \sqrt{m_A E_r / 2}$ is the minimum neutrino energy required to produce a NR energy $E_r$. The total expected event rate from the DSNB components (all six flavors) integrated over recoil energies of 0--10~keV$_{\rm nr}$ (NR energy) is approximately 0.025~events/tonne/year.

The top panel of Fig.~\ref{fig:flux_and_rate} illustrates the incident neutrino flux spectra for the comparison of DSNB and the irreducible atmospheric and solar neutrino backgrounds due to the same experimental signature, and the bottom panel of  Fig.~\ref{fig:flux_and_rate} shows their corresponding expected CE$\nu$NS event rate as a function of NR energy in the detector. 

\textit{Detector}.
The PandaX-4T experiment is located in the B2 hall of CJPL-$\textup{\uppercase\expandafter{\romannumeral 2}}$ \cite{KJKang2010}, with 2.4 km rock overburden. The cosmic muon flux at CJPL is measured to be $\mathcal{O}(10^{-10}) \text{ cm}^{-2}\text{ s}^{-1}$ \cite{Guo2021}. The detector utilizes a two-phase xenon time projection chamber (TPC), with an active target mass of 3.7 tonnes of liquid xenon in its sensitive region. Energy deposition in the liquid xenon produces prompt scintillation photons (${S1}$) via atomic de-excitation and electron-ion recombination, while ionized electrons drift to the gaseous phase, generating delayed electroluminescence photons (${S2}$) through proportional scintillation.

\textit{Reconstruction}.  
Arrays of three-inch Hamamatsu R11410-23 photomultiplier tubes (PMTs) at the top and bottom regions of the TPC capture photons from both ${S1}$ and ${S2}$ signals, enabling 3D vertex reconstruction through the hit pattern and ${S1–S2}$ time delay. The electron-equivalent recoil energy ${E_{ee}}$ of a given event is reconstructed as~\cite{MSzydagis_2011}
\begin{equation}
    E_{ee} = W \left( \frac{Q_{S1}^{c}}{g_{1}} + \frac{Q_{S2_{b}}^{c}}{g_{2_b}} \right),
\end{equation}
where $W=13.7$ eV is the mean energy to produce a single photon or an ionized electron in liquid xenon. $Q_{S1}^{c}$ is the corrected photo-electron (PE) charges for the total ${S1}$, and $Q_{S2_{b}}^{c}$ is the corrected ${S2}$ charges detected by the bottom PMT arrays. $g1$ is the photon detection efficiency for ${S1}$, and $g_{2_{b}}$ combines the electron extraction efficiency with the single-electron gain of the bottom-array PMTs. 



\textit{Data \& selection}.  
In this analysis, we utilize the data from Run0 and Run1, which are divided into several subsets based on different electron lifetimes, liquid levels, cathode voltages, and gate voltages, as detailed in \cite{PRL_NewPysics}. Data quality filters are the same as Ref~\cite{PhysRevLett2024}. 
Due to the self-shielding capability of liquid xenon, external material backgrounds decrease toward the center of the detector, whereas the DSNB signal is expected to be uniformly distributed throughout the TPC. The fiducial masses are determined to be 2.38~$\pm$~0.04~tonnes for Run0 and 2.48~$\pm$~0.05~tonnes for Run1~\cite{PhysRevLett2024}. Following the standard PandaX-4T analysis~\cite{PhysRevD023029}, additional event selection criteria are applied, including accidental coincidence (AC) rejection, $S1$ and $S2$ quality cuts, and requirements for a single $S1$ and a single $S2$ signal, etc.  


Defining the lower boundaries of region of interest (ROI) is crucial, given the sharp rise of ACs and solar $^{8}\mathrm{B}$ events at the lower energy region. We perform an exhaustive four-dimensional grid search for the lower and upper bounds of both the ${S1}$ and $S2_b$ signals to maximize the expected statistical significance, which is evaluated using Cowan's formula \cite{cowan2012discovery},  accounting for the systematic uncertainty of the background: 

\begin{equation}
\begin{split}
Z = \Biggl[ 2 \biggl( & (s+b) \ln \left[ \frac{(s+b)(b+\sigma_b^2)}{b^2+(s+b)\sigma_b^2} \right] \\
& - \frac{b^2}{\sigma_b^2} \ln \left[ 1+\frac{\sigma_b^2 s}{b(b+\sigma_b^2)} \right] \biggr) \Biggr] ^{1/2},
\end{split}
\end{equation}
where $s$ and $b$ are the expected signal and background counts, and $\sigma_b$ represents the systematic uncertainty of the background. The background model will be discussed later. Based on the optimization formula, the ${S1}$ signal is required to have a minimum coincidence of two fired PMTs. However, events with ${S1}$ below 3 PEs are excluded from the optimization space because the AC background becomes overwhelmingly high at such a low threshold, severely degrading the significance. The optimized ROIs are $S1=3$--$13$~PEs and $S2_b=50$--$225$~PEs for Run0, and $S1=3$--$15$~PEs and $S2_b=70$--$335$~PEs for Run1. In both runs, the DSNB search ROI lies entirely within the corresponding dark-matter search ROI reported in Ref.~\cite{PhysRevLett2024}.

\begin{figure}[htbp]
    \centering
    \includegraphics[width=0.48\textwidth]{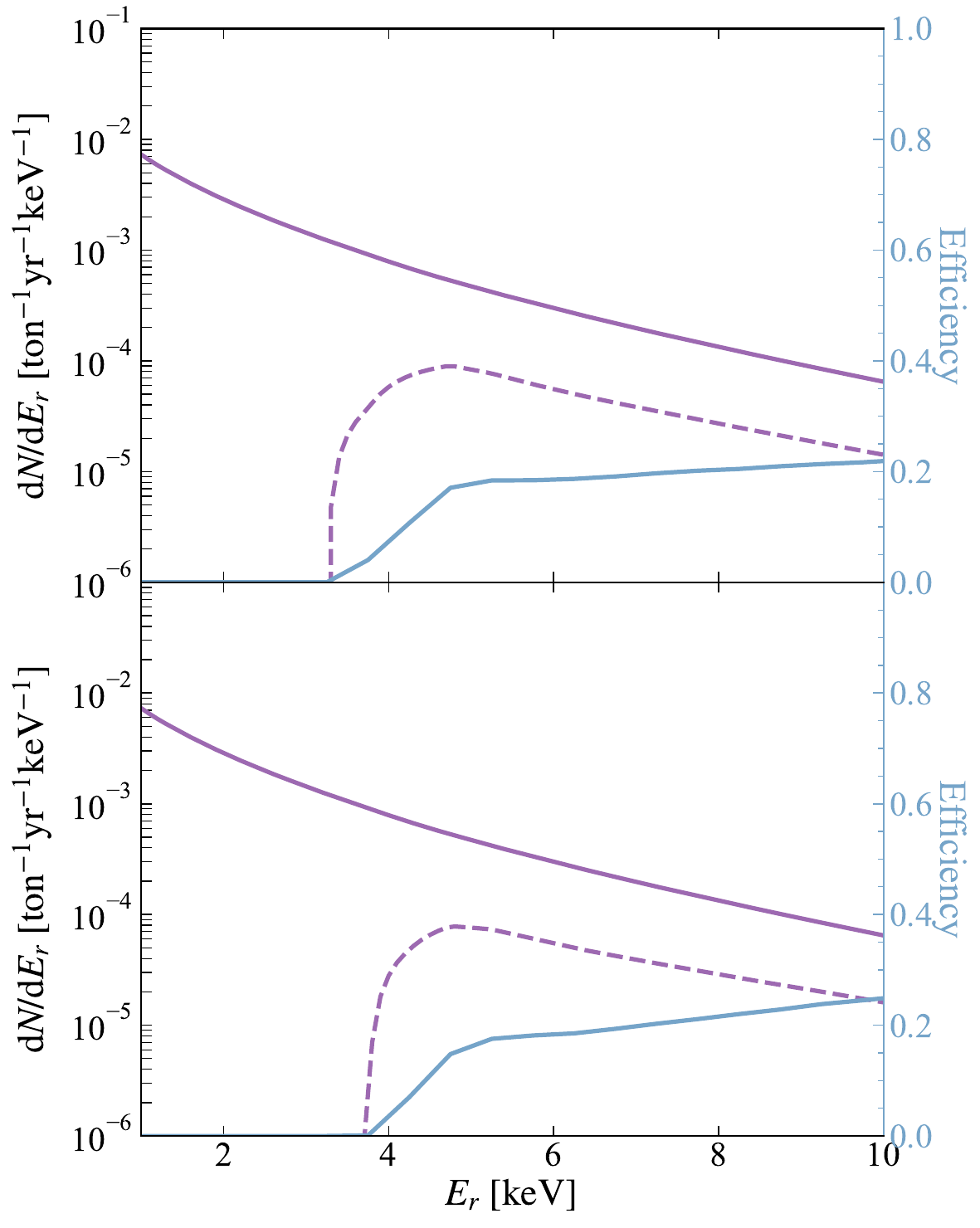}
    \caption{Energy spectra and detection efficiencies for Run0 (top) and Run1 (bottom). The solid and dashed purple curves represent the theoretical and the expected event rate spectra before and after the efficiency correction, respectively, while the solid blue curves denote the detection efficiencies.} 
    \label{fig:efficiency_curve}
\end{figure}

The energy-dependent DSNB detection efficiency, illustrated in Fig.~\ref{fig:efficiency_curve}, is evaluated utilizing a combination of data-driven techniques and waveform simulations. It can be factorized into three primary components: the detection efficiency after basic cuts~\cite{PhysRevD023029} ($\sim$4\%), the acceptance of the optimized ${S1}$--${S2_b}$ ROI (74\% and 71\% for Run0 and Run1), and the signal survival probability after the below-NR-median cut (70\% for Run0 and 72\% for Run1). The detection efficiency after basic cuts is low, because DSNB NR events tend to cluster at low energies and therefore produce no visible observable above the detector threshold. Convolving the total efficiency with the target mass and effective exposure yields an expected DSNB signal of $(3 \pm 1) \times 10^{-4}$ events in Run0 and $(5 \pm 3) \times 10^{-4}$ events in Run1.  Given the low predicted signal count, the observed events will be background-dominated. 

\textit{Backgrounds}.  
The primary background estimation follows our previous work~\cite{PhysRevLett2024}, with the addition of two new components specific to this search: solar Hep neutrinos and atmospheric neutrinos. The backgrounds within the ROI can be classified into three distinct categories: (1) NR backgrounds, primarily from neutrons and other neutrino sources; (2) electron recoil (ER) backgrounds, originating from $\beta$ and $\gamma$ decays of radioactive isotopes; and (3) unphysical backgrounds, dominant by accidental coincidences of isolated signals. The nominal expected event counts and the best-fit results from a background-only fit are summarized for all components in Table~\ref{tab:background_summary}.  These background estimations in the ROI are detailed as follows:

\begin{table*}[htbp]
  \caption{Summary of expected background events within the optimized ROIs: Run0 ($S1\in[3,13]$~PE, $S2_b\in[50,225]$~PE) and Run1 ($S1\in[3,15]$~PE, $S2_b\in[70,335]$~PE). Also shown are the best-fit values from the likelihood fit and the observed event counts.}
  \label{tab:background_summary}
  \centering
  \begin{tabular*}{0.9\textwidth}{@{\extracolsep{\fill}}clcccc@{}}
    \toprule
    \multirow{2}{*}{Category} & \multirow{2}{*}{Background} & \multicolumn{2}{c}{Run0} & \multicolumn{2}{c}{Run1} \\
    \cmidrule(lr){3-4} \cmidrule(lr){5-6}
    & & Expected & Fitted & Expected & Fitted \\
    \midrule
    
    \multirow{8}{*}{ER} 
    & $^{214}\mathrm{Pb}$          & $0.0134 \pm 0.0005$  & $0.0134 \pm 0.0012$ & $0.0473 \pm 0.0019$  & $0.0469^{+0.0057}_{-0.0056}$ \\
    & $^{212}\mathrm{Pb}$          & $0.0029 \pm 0.0008$  & $0.0029 \pm 0.0011$ & $0.0072 \pm 0.0019$  & $0.0071 \pm 0.0028$ \\
    & $^{85}\mathrm{Kr}$           & $0.0047 \pm 0.0024$  & $0.0047 \pm 0.0035$ & $0.0198 \pm 0.0057$  & $0.0197 \pm 0.0100$ \\
    & $\mathrm{pp} + {}^{7}\mathrm{Be}$ & $0.0018 \pm 0.0002$ & $0.0018 \pm 0.0003$ & $0.0041 \pm 0.0004$  & $0.0041 \pm 0.0007$ \\
    & Tritium                      & $0.0787 \pm 0.0048$  & $0.0788 \pm 0.0087$ & $0.0252 \pm 0.0073$  & $0.0250 \pm 0.0050$ \\
    & $^{136}\mathrm{Xe}$          & $0.0002 \pm 0.0000$  & $0.0002 \pm 0.0000$ & $0.0008 \pm 0.0000$  & $0.0008 \pm 0.0001$ \\
    & Surface                      & $0.0262 \pm 0.0170$  & $0.0257 \pm 0.0242$ & $0.0372 \pm 0.0242$  & $0.0361 \pm 0.0341$ \\ 
    & Material ER                  & $0.0023 \pm 0.0003$  & $0.0023 \pm 0.0004$ & $0.0064 \pm 0.0007$  & $0.0063 \pm 0.0012$ \\

    \midrule
    
    \multirow{4}{*}{NR} 
    & Neutron                      & $0.1925 \pm 0.0427$  & $0.1684 \pm 0.0507$ & $0.3023 \pm 0.0462$  & $0.2513^{+0.0522}_{-0.0520}$ \\
    & $^{8}\mathrm{B}$ $\nu$       & $0.1123 \pm 0.0348$  & $0.1130^{+0.0500}_{-0.0496}$ & $0.1352 \pm 0.0419$  & $0.1349^{+0.0607}_{-0.0601}$ \\
    & Hep $\nu$                    & $0.0045 \pm 0.0014$  & $0.0045 \pm 0.0019$ & $0.0070 \pm 0.0021$  & $0.0069 \pm 0.0030$ \\
    & Atmospheric $\nu$            & $0.0015 \pm 0.0004$  & $0.0015 \pm 0.0005$ & $0.0036 \pm 0.0009$  & $0.0036 \pm 0.0013$ \\
    \midrule
    

    Unphysical 
    & Accidental                   & $1.4772 \pm 0.4431$  & $1.3679^{+0.5871}_{-0.5728}$ & $1.9414 \pm 0.5824$  & $1.7825^{+0.7785}_{-0.7597}$ \\ 
    \midrule  
    
    \multicolumn{2}{l}{Total}      & $1.9183 \pm 0.4469$  & $1.7851^{+0.5920}_{-0.5778}$ & $2.5374 \pm 0.5863$  & $2.3251^{+0.7835}_{-0.7647}$ \\
    \multicolumn{2}{l}{Observed}   & \multicolumn{1}{c}{2} & $\cdot\cdot\cdot$ & \multicolumn{1}{c}{1} & $\cdot\cdot\cdot$ \\
    \bottomrule
  \end{tabular*}
\end{table*}


The neutron background originates from $(\alpha, n)$ reactions and the spontaneous fission of long-lived isotopes (e.g., $^{232}\mathrm{Th}$, $^{235}\mathrm{U}$, $^{238}\mathrm{U}$) within the detector materials. 
Following Ref.~\cite{Huang_2022}, the expected contributions from neutron backgrounds are determined to be 0.1925 events for Run0 and 0.3023 events for Run1. 

The CE$\nu$NS events from other neutrino sources constitutes an irreducible NR background. At lower energies, solar neutrinos ($^{8}\mathrm{B}$ and $hep$) dominate, while atmospheric neutrinos prevail at higher energies. For the $^{8}\mathrm{B}$ neutrino background estimation, we adopt the same method as Ref~\cite{PhysRevLett2024}. For the atmospheric neutrino background, we adopt a site-specific, solar-cycle-averaged flux at CJPL~\cite{CJPL_atmospheric,Fan_atmospheric}. Taking into account systematic uncertainties of 25\% for atmospheric \cite{Fan_atmospheric} and 30\% for $hep$ neutrinos \cite{Vinyoles_2017}, the total expected CE$\nu$NS background is 0.1183 events for Run0 and 0.1458 events for Run1 (see Table~\ref{tab:background_summary} for break-down). 


Radon isotopes ($^{222}\mathrm{Rn}$ and $^{220}\mathrm{Rn}$) emanating from detector materials and cryogenic circulation pipelines contribute to the low-energy ER background primarily via the $\beta$-decays of their progenies, $^{214}\mathrm{Pb}$ and $^{212}\mathrm{Pb}$. 
Following Ref.~\cite{Ma:2020kll, PhysRevLettXe134}, the expected $^{214}\mathrm{Pb}$ background yields are 0.0134 events for Run0 and 0.0473 events for Run1 within the ROI. And the expected $^{212}\mathrm{Pb}$ background yields are 0.0029 events for Run0 and 0.0072 events for Run1.  

The $^{85}\mathrm{Kr}$ $\beta$-decay contribution is evaluated using the delayed $\beta$-$\gamma$ coincidence from $^{85}\mathrm{Kr}$ to $^{85}\mathrm{Rb}$. Based on the measured average Kr/Xe concentrations ($0.52 \pm 0.27$~ppt for Run0, $0.94 \pm 0.28$~ppt for Run1) and assuming an isotopic abundance of $2 \times 10^{-11}$ \cite{Collon_2004}, the expected counts are 0.0047 and 0.0198 events, respectively.

The tritium coming from the PandaX-II end-of-run calibration contributes via low-energy $\beta$ decays. The rate is extracted by fitting the one-dimensional $S1$ spectrum, giving the average rate $2.04 \pm 0.25$~events/tonne/day in Run0. Following online distillation \cite{PhysRevLett2024}, this rate was significantly reduced to $0.26 \pm 0.07$~events/tonne/day in Run1. The expected events are 0.0787 for Run0 and 0.0252 for Run1 within the ROI.

The solar neutrinos ($pp$ and ${}^{7}\mathrm{Be}$) elastically scatter off electrons. With standard solar neutrino fluxes and three-flavor oscillations \cite{pandaxcollaboration2024measurementsolarppneutrino}, they contribute 0.0018 and 0.0041 events in Run0 and Run1. 

The xenon isotope $^{136}\mathrm{Xe}$ undergoes two-neutrino double-beta decay within the detector volume. Based on its measured isotopic abundances of $(8.86\pm0.17)\%$~\cite{Xe136}, the expected event counts are 0.0002 and 0.0008 in Run0 and Run1. 

Material backgrounds arise from $\beta$ and $\gamma$ emissions of isotopes intrinsic to the detector components, primarily $^{60}$Co, $^{137}$Cs, $^{40}$K, and the $^{232}$Th, $^{235}$U, and $^{238}$U decay chains. These ER contributions are evaluated using the Geant4-based Monte Carlo framework BambooMC~\cite{Chen_2021}, which predicts 0.0023 and 0.0064 events for Run0 and Run1.

Surface backgrounds mainly originate from $\beta$-decays of $^{210}\mathrm{Pb}$ accumulated on the PTFE inner walls. Electric field distortion near the detector borders causes charge absorption, reducing ${S2}$ signals and leading to their misclassification as NR events. We model this charge loss using ${S2}$ spectra from $^{210}\mathrm{Po}$ $\alpha$-decays to construct a 3D spatial distribution within the ${S1}$--${S2_b}$--radius parameter space~\cite{QIAN2025170548}. This model predicts a total background contribution of 0.0262 events for Run0 and 0.0372 events for Run1.


The AC background arises from the random pairing of physically uncorrelated isolated ${S1}$ and ${S2}$ signals within the maximum drift time window~\cite{Abdukerim_2022}. The expected AC rate is modeled using data-driven rates of isolated ${S1}$ (13--14~Hz) and ${S2}$ (0.13--0.15~Hz) signals. This analytical model is cross-checked against an independent off-window measurement, which samples unphysical events with drift times between 860 and 1500~$\mu$s. The two estimations are consistent with each other, with the difference as the systematic uncertainty. As the dominant background in the ROI, the AC contribution is estimated to be 1.4772 events for Run0 and 1.9414 events for Run1.


We validate the background model by comparing observed and predicted event counts in higher-energy sideband regions, which are defined as ${S1} \in [20, 30]$~PEs for both datasets, with ${S2_b} \in [350, 500]$~PEs for Run0 and $[550, 700]$~PEs for Run1. We observe one event in Run0 (expected: $1.19 \pm 0.07$) and one event in Run1 (expected: $1.35 \pm 0.11$). These results are fully consistent with statistical fluctuations, confirming the model's reliability for signal ROI extrapolation.  

\textit{Data interpretation \& results}.   
The statistical analysis employs a two-dimensional unbinned extended likelihood function in the $(S1,\, \log_{10}(S2_b/S1))$ parameter space. The likelihood function is constructed as:

\begin{equation*}
\begin{split}
	\mathcal{L} = & \left [ \prod_{n=1}^{n_{\rm set}=11}\mathcal{L}_{n}\right ] \times \left [ \prod_{b}G(\delta_{b},\sigma_{b} ) \right ] \times \left [ \prod_{\epsilon}G(\delta_{\epsilon},\sigma_{\epsilon} ) \right ] 
\end{split}
\end{equation*}
with
\begin{equation}
\begin{split}
	\mathcal{L}_n = & {\rm Poisson}(n_{\rm obs}^n \mid n_{\rm fit}^n) 
	\times \Biggl[ \prod_{i=1}^{n_{\rm obs}^n} \frac{1}{n_{\rm fit}^n} \biggl( N_s^n P_s^n\big(S1^i, S2_b^i) \\
	& + \sum_b N_b^n(1+\delta_b) P_b^n\big(S1^i, S2_b^i \big) \biggr) \Biggr]
\end{split}
\end{equation}
\begin{equation}
	G(\delta_b,\sigma_b) = \frac{1}{\sqrt{2\pi}\sigma_b} \exp\left(-\frac{\delta_b^{2}}{2\sigma _b^{2}} \right)
\end{equation}

\begin{equation}
	G(\delta_{\epsilon},\sigma_{\epsilon}) = \frac{1}{\sqrt{2\pi}\sigma_{\epsilon}} \exp\left(-\frac{\delta_{\epsilon}^{2}}{2\sigma _{\epsilon}^{2}} \right)
\end{equation}

Here, $n_{\rm set}$ denotes the total number of analysis subsets, with $n_{\rm obs}^n$ being the observed event count in subset $n$. Both the nominal signal and background expectations ($N_{s,\rm exp}^n$ and $N_{b,\rm exp}^n$) are scaled by the global efficiency nuisance parameter $\delta_{\epsilon}$ to obtain the signal count $N_s^n$ and background count $N_b^n$, respectively. The total fitted count is defined as  $n_{\rm fit}^n = N_s^n + \sum\limits_b N_b^n(1+\delta_b)$, where $\delta_b$ captures the expected yield uncertainty specific to each background component $b$. 

Systematic uncertainties are incorporated into the likelihood via Gaussian penalty terms. The normalization parameters ($\delta_{\epsilon}$ and $\delta_b$) are constrained by one-dimensional Gaussians $G(\delta, \sigma)$. The relative uncertainties for signal efficiency and various background components are detailed in Table~\ref{tab:systematic_uncertainties}. The efficiency uncertainty ($\sigma_{\epsilon}$) is derived from calibration and waveform simulations. 

\begin{table}[htbp]
  \caption{Summary of systematic uncertainties for signal efficiency and background normalization parameters. Uncertainties that are constant across all datasets are denoted as ``All'' in the Run/Set column, whereas run-dependent or set-dependent values (e.g., Tritium in Run0) are listed explicitly for their corresponding data partitions.}
  \label{tab:systematic_uncertainties}
  \centering
  \begin{tabular}{lllc}
    \toprule
    Category & Source & Run / Set & Uncertainty ($\sigma$) \\
    \midrule
    
    \multirow{2}{*}{Efficiency} 
    & \multirow{2}{*}{Signal Efficiency} & Run0 & 0.05 \\
    &                                    & Run1 & 0.075 \\
    \midrule
    
    \multirow{23}{*}{Background}
    & \multirow{2}{*}{$^{85}\mathrm{Kr}$} & Run0 & 0.52 \\
    &                                     & Run1 & 0.29 \\
    \cmidrule{2-4}
    & \multirow{2}{*}{Neutron}     & Run0 & 0.30 \\
    &                                     & Run1 & 0.21 \\
    \cmidrule{2-4}
    & \multirow{6}{*}{Tritium}            & Run0, Set 1 & 0.33 \\
    &                                     & Run0, Set 2 & 0.13 \\
    &                                     & Run0, Set 3 & 0.73 \\
    &                                     & Run0, Set 4 & 0.08 \\
    &                                     & Run0, Set 5 & 0.12 \\
    &                                     & Run1 & 0.29 \\
    \cmidrule{2-4}
    & $^{214}\mathrm{Pb}$                 & All & 0.04 \\
    & $^{136}\mathrm{Xe}$                 & All & 0.05 \\
    & $\mathrm{pp} + {}^{7}\mathrm{Be}$   & All & 0.10 \\
    & Material                            & All & 0.11 \\
    & Atmospheric $\nu$                   & All & 0.25 \\
    & $^{212}\mathrm{Pb}$                 & All & 0.27 \\
    & Hep $\nu$                           & All & 0.30 \\
    & Accidental                          & All & 0.30 \\
    & $^{8}\mathrm{B}$ $\nu$              & All & 0.31 \\
    & Surface                             & All & 0.65 \\
    \bottomrule
  \end{tabular}
\end{table}

The probability density functions (PDFs), $P_s^n(S1, S2_b)$ and $P_b^n(S1, S2_b)$, describe the probability distributions of the measured $(S1, S2_b)$ observables for the signal and type-$b$ background components, respectively, in subset $n$. These PDFs are generated using the NEST (Noble Element Simulation Technique) model~\cite{PhysRevD023029}, which self-consistently describes the liquid xenon scintillation and ionization microphysics.

\begin{figure}[htbp]
    \centering
    \includegraphics[width=0.48\textwidth]{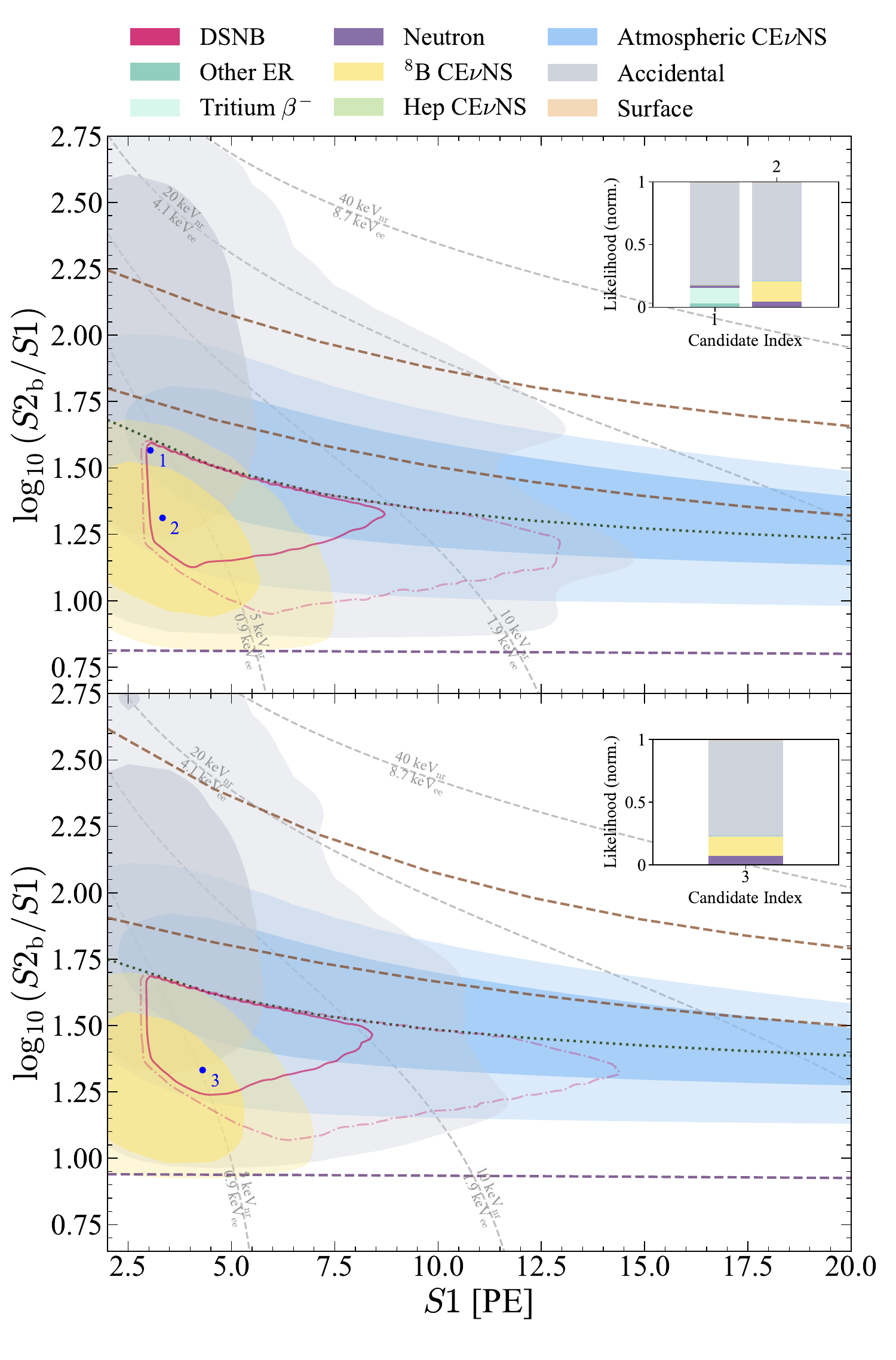}
    \caption{Event distributions in $(S1, \mathrm{log_{10}}(S2_b/S1))$ parameter space for Run0 (top panel) and Run1 (bottom panel). The 1$\sigma$ and 2$\sigma$ search regions for the DSNB signal are defined by the solid and dashed pink contours. The darker and lighter shaded bands illustrate the 1$\sigma$ and 2$\sigma$ regions for various background components, including accidental coincidences (grey), $^8$B CE$\nu$NS (yellow), and atmospheric CE$\nu$NS (blue). Additional reference curves are overlaid: the 5\% and 95\% ER quantiles (dark brown dashed), the NR median (green dotted), and the 99.5\% NR acceptance boundaries (purple dashed). Three candidates falling within the signal region are labeled with blue numbers (1 and 2 in Run0; 3 in Run1). The inset bar plots show the signal and background probabilities for these candidates.}
    \label{fig:unblinded_data}
\end{figure}

In the ROI, we observe 2 events in Run0, and 1 event in Run1, which are consistent with the expected background estimation. The unbinned maximum likelihood fit is performed on the combined dataset, yielding a best-fit number of null DSNB signal events. The observed events in the $(S1, \log_{10}(S2_b/S1))$ parameter space, along with the background and signal fittings, are shown in Fig.~\ref{fig:unblinded_data}. Each event candidate is characterized by the likelihoods of different background components.

Based on the above, we proceed to place physical constraints on the DSNB $\nu_x$ flux. The 90\% C.L. upper limit is derived utilizing the profile likelihood ratio (PLR)~\cite{Baxter:2021pqo}, and the upper limit on the DSNB flux, $\Phi_{90}$, is derived from the 90\% C.L. upper limit on the number of signal events, $N_{90}$, via the scaling relation:

\begin{equation}
  \Phi_{90} = \frac{N_{90}}{N_{\text{exp}}} \times \Phi_{\text{model}}, 
\end{equation}
where $N_{\text{exp}} = \text{0.0008}$ is the expected number of signal events that can be detected in the PandaX-4T in this analysis after all of the selection cuts, with the corresponding theoretical $\nu_x$ (each of $\nu_\mu$, $\nu_\tau$, $\bar{\nu}_\mu$, $\bar{\nu}_\tau$) flux of  DSNB model calculated to be $\Phi_{\text{model}} = \text{0.67} \text{ cm}^{-2}\text{s}^{-1}$ for incident neutrino energies exceeding 19.3~MeV.  Applying our calculated upper limit of $N_{90} = \text{2.71}$, we obtain a 90\% C.L. upper limit of $\Phi_{90} = 2.27 \times 10^3 \text{ cm}^{-2}\text{s}^{-1}$ for incident neutrino energies exceeding 19.3~MeV. Taking into account the DSNB flux uncertainties shown by the band in the top panel of Fig.~\ref{fig:flux_and_rate}, the upper limit is calculated to be $\Phi_{90} = (2.27 - 2.43) \times 10^3 \text{ cm}^{-2}\text{s}^{-1}$.  The result is comparable to the existing world best limit, which is $\Phi_{90}$ = $ (1.0 - 1.4) \times 10^3 \text{ cm}^{-2}\text{s}^{-1}$ for $\nu_\mu$ and $\nu_\tau$, and $ (1.3 - 1.8) \times 10^3 \text{ cm}^{-2}\text{s}^{-1}$ for $\bar{\nu}_\mu$ and $\bar{\nu}_\tau$, based on an exposure of $92.2~\mathrm{kt}\cdot\mathrm{year}$,  as reported in Ref.~\cite{Lunardini_2008}. Additionally, a recent LZ preprint reports a more stringent limit of $\Phi_{90} = (686-826)~\mathrm{cm}^{-2}\mathrm{s}^{-1}$~\cite{Xia:2024ysq}. This result is not yet peer reviewed. 


\textit{Summary}. 
In summary, we report the first experimental search for the DSNB via CE$\nu$NS with the PandaX-4T detector at the CJPL. Based on an exposure of $1.54~\mathrm{tonne}\cdot\mathrm{year}$ from the commissioning and first science runs, we observe no statistically significant event excess above the background expectation. Therefore, we place a 90\% C.L. upper limit of $(2.27- 2.43) \times 10^3~\mathrm{cm^{-2}s^{-1}}$ on the DSNB $\nu_x$ flux per flavor, for neutrino energies $E_\nu > 19.3$~MeV. The result is comparable to the existing world best limit, despite having an exposure four to five orders of magnitude smaller than that of SK. With the future PandaX-20T experiment, we anticipate a significant enhancement of this limit, offering a promising path toward probing the $\nu_x$-flavor DSNB with higher sensitivity. 

\textit{Acknowledgments}. 
 This project is supported in part by grants from National Key R\&D Program of China (Nos. 2023YFA1606200, 2023YFA1606203), National Science Foundation of China (Nos. 12090060, 12090061, U23B2070), and by Office of Science and Technology, Shanghai Municipal Government (grant Nos. 21TQ1400218, 22JC1410100, 23JC1410200, ZJ2023-ZD-003). We thank for the support by the Discipline Construction Fund of Shandong University, and the Fundamental Research Funds for the Central Universities. We also thank the sponsorship from the Chinese Academy of Sciences Center for Excellence in Particle Physics (CCEPP), Thomas and Linda Lau Family Foundation, New Cornerstone Science Foundation, Tencent Foundation in China, and Yangyang Development Fund. Finally, we thank the CJPL administration and the Yalong River Hydropower Development Company Ltd. for indispensable logistical support and other help.

\bibliography{refs.bib}
\end{document}